\documentclass[a4paper,10pt]{article}

\usepackage{epsfig}
\usepackage{hyperref}
\usepackage{amsmath, amsthm, amssymb}
\include{qcircuit}
\newcommand{\inprod}[2]{\langle #1 | #2 \rangle}
\newcommand{\oper}[2]{| #1 \rangle \langle #2 |}
\def\bp{\begin{array}{c} \begin{picture}}
\def\ep{\end{picture}\end{array}}

\newtheorem{thm}{Theorem}
\newtheorem{lem}{Lemma}
\newtheorem{defn}{Definition}

\begin{document}

\thispagestyle{empty}
\setcounter{page}{1}

\vspace*{0.88truein}

\centerline{\bf CLUSTER STATE QUANTUM COMPUTATION}
\vspace*{0.035truein}
\centerline{\bf FOR MANY-LEVEL SYSTEMS}
\vspace*{0.37truein}
\centerline{\footnotesize WILLIAM HALL}
\vspace*{0.015truein}
\centerline{\footnotesize\it Department of Mathematics, University of York}
\centerline{\footnotesize\it Heslington, York YO10 5DD, United Kingdom}
\vspace*{0.225truein}

\vspace*{0.21truein}

\begin{abstract}
The cluster state model for quantum computation [Phys. Rev. Lett. \textbf{86}, 5188] outlines a scheme that allows one to use measurement on a large set of entangled quantum systems in what is known as a cluster state to undertake quantum computations. The model itself and many works dedicated to it involve using entangled qubits. In this paper we consider the issue of using entangled qudits instead. We present a complete framework for cluster state quantum computation using qudits, which not only contains the features of the original qubit model but also contains the new idea of adaptive computation: via a change in the classical computation that helps to correct the errors that are inherent in the model, the implemented quantum computation can be changed. This feature arises through the extra degrees of freedom that appear when using qudits. Finally, for prime dimensions, we give a very explicit description of the model, making use of mutually unbiased bases.
\end{abstract}

\vspace*{10pt}

\vspace*{3pt}

\vspace*{1pt}  
\section{Introduction}

The problem of building a quantum computer that has computational power greater than that of classical computers is one that has concerned both theorists and experimentalists for a number of years. The recent development of the cluster state model for quantum computation \cite{CS1, CS2, CS3}, in which remarkably quantum computation is achieved essentially by tailored measurement of entangled states of a large number of qubits, has not only lead to significant implications in our understanding of quantum computation and quantum dynamics, but also to new experimental approaches to the possible realisation of large scale quantum computers \cite{CS_O, CS_O2, CS_E}. 

Although much study has been devoted to this model \cite{CS_VB, CS_F, CS_shape, LCS, CS_F2} this work has almost exclusively been concerned with using qubits as the basic physical resource. However, many quantum systems cannot be treated as simple two-level systems, but rather as multi-level systems. Furthermore, recent work \cite{QCHSM} has suggested that qutrit based quantum computation schemes yield in some sense the most efficient implementation of quantum computation. 

In this paper we are going to present a generalisation of the cluster state model, by replacing the cluster state of qubits with more general cluster states of qudits.  Such a generalisation has already been presented in \cite{dCQC}, but in this paper we will take a much more explicit approach, in which we try and present the cluster state model by presenting each constituent of the model in as simple a way possible. This allows us to see the critical ingredients that make cluster state QC possible.

The paper will be laid out as follows. In section \ref{sec_1dt} we will present a generalisaton of an idea used in \cite{NJC} which we will call \emph{one dit teleportation}. This forms the core to our approach to cluster state QC. One dit teleportation takes the form of a simple two-qudit circuit identity which can be thought of as implementing a certain quantum gate (namely the $d$-dimensional analogue of the Fourier transform gate) by measurement. We then proceed to show that this identity can be used to implement a much wider class of gates, and how this process can be turned into a quantum computational paradigm, which essentially involves a series of measurements on a entangled state of qudits to implement quantum computation. While this approach turns out to not be the best approach for some types of cluster states, it ties in our cluster state model for QC  heavily with the circuit model, hence giving us an alternative and (in the eyes of the author) a simpler way of looking at the model.

In section \ref{sec_ad} we will first discuss the issue of adapting measurements based on previous measurement results to allow us to implement any quantum operation we wish. Measurements on a quantum system by their very nature are statistical; and in the cluster state paradigm this randomness can be thought of as introducing errors into our computation which can be corrected for using \emph{classical} computation. This notion is known as the issue of \emph{adaptive measurement} and is crucial in the theory of cluster state QC. We will also introduce a new phenomena which we call \emph{adaptive computation}, a feature that will be unique to cluster state QC on qudits, arising due to the extra degrees of freedom we are presented with. The basic idea is that by changing the method by which we correct the errors associated with a measurement outcome, we can implement different quantum gates. We will discuss this in further detail at the appropriate time. We conclude this section with an example of the simplicity of the one-dit teleportation approach which also makes use of adaptive computation.

In section \ref{sec_stab}, we briefly review the stabiliser method for cluster states, which is the most common approach in most of the existing literature. For details of the model we refer the reader to the appropriate paper, but we do give a comparison of the two approaches, highlighting the fact that both approaches are indeed useful.

In section \ref{sec_ung} we will discuss the problem of generating a universal set of quantum gates for quantum computation, giving a general method by which quantum gates can be realised, and then discussing a connection of this method with the theory of \emph{mutually unbiased bases}. While this connection is not a fundamental one, using the known theory for MUBs gives us a very elegant method of establishing measurement patterns for implementing a universal set of quantum gates. We will use the connection to give explicit calculations and details of the measurements required to generate certain quantum gates in prime dimensions. We conclude with a discussion in the final section.

\section{One dit teleportation and cluster state quantum computation} \label{sec_1dt}

\subsection{Preliminaries: a basic measurement result}

For clarity, we will start by giving a very basic result. We assume throughout this paper that all Hilbert space(s) involved here are of dimension $d$. Let $\{ \ket{k} \}_{k=0}^{d-1} $ be a standard (computational) basis for our Hilbert space. We define \emph{measuring in the basis defined by $U$} for a unitary $U \in U(d)$ to mean measurement in the orthonormal basis $\{U\ket{k} \}_{k=0}^{d-1} $ (i.e. the vectors defined by the columns of $U$ when $U$ is described in matrix form in the computational basis). Our result is as follows:
\begin{lem} Given a bipartite state $\ket{\Psi}$, measuring one of the systems in the basis defined by $U$ and discarding is equivalent to applying $U^\dagger$ to the first system, measuring it in the computational basis and then discarding it. \label{M} \end{lem}
\proof{ Let $\ket{\Psi} = \sum_k U\ket{k} \otimes \ket{\psi_k}$. Measuring the first system in the basis defined by $U$ 
leaves the second system in the state $\ket{\psi_k}/ \sqrt{\inprod{\psi_k}{\psi_k}}$ with probability $\inprod{\psi_k}{\psi_k}$. Applying $U^\dagger$ to the first system and then measuring it in the computational basis can easily be seen to have the same effect. }

This lemma gives us an alternative way to generate the appropriate statistics for measurement of one system of a bipartite system in a given basis.

\subsection{One dit teleportation}

The idea of one dit teleportation is a many-level generalisation of \emph{one-bit teleportation} that was first presented in \cite{1BT}. We will first simply present this relatively simple idea, and then show how it is integral to cluster state quantum compuation.

We first need some definitions. The Fourier transform basis is defined by 
\begin{equation} \ket{+_j} = \sum_{k=0}^{d-1} \omega^{jk} \ket{k} \label{Fb} \end{equation}
where $\omega=e^{2 \pi i / d}$, the primitive $d$th root of unity. We will often denote $\ket{+_0}$ by simply $\ket{+}$. We also need to introduce a set of operators known as the \emph{generalised Pauli operators} \cite{GP1}: 
\begin{equation} Z = \sum_{k=0}^{d-1} \omega^k \oper{k}{k}, \ X = \sum_{k=0}^{d-1} \oper{k-1}{k} \end{equation}
where we are using modulo $d$ arithmetic within the bras and kets. Finally, we define the Fourier gate $F$ by
\begin{equation} F = \frac{1}{\sqrt{d}} \sum_{j,k=0}^{d-1} \omega^{jk} \oper{j}{k} = \sum_{k=0}^{d-1} \oper{+_k}{k}. \end{equation}
This gate is the $d$-dimensional Quantum Fourier Transform gate, and in $d=2$ is equivalent to the Hadamard gate.

\begin{figure}[!htp] 
\[ 
\Qcircuit @C=0.4em @R=0.5em @!R{
\lstick{\ket{\psi}} & \ctrl{1} & \qw & \gate{F} & \meter & \cw & \rstick{m} \\
\lstick{\ket{+}} & \gate{Z} & \qw & \qw & \qw & \qw & \rstick{X^m F \ket{\psi}} } 
\]
\caption{The circuit diagram for one-dit teleportation. \label{t}}
\end{figure}
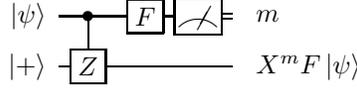

The notion of one-dit teleportation is given by the circuit identity in Figure \ref{t}, where we are using the $d$ dimensional analog of controlled gates:
\begin{equation}  CZ = \sum_{k=0}^{d-1} \oper{k}{k} \otimes Z^k = \sum_{k,l=0}^{d-1} e^{2\pi i kl/d} \oper{k}{k} \otimes \oper{l}{l} \label{CZ} \end{equation}
and the meter represents measurement in the computational basis, with outcome $k$ corresponding to the state $\ket{k}$. Furthermore, each of the possible outcomes occurs with equal probability $1/d$.

\proof{First we note two important facts, both of which can be deduced by elementary means:
\begin{equation} Z^j \ket{+_k} = \ket{+_{j+k}} \textrm{ (modulo $d$ addition in ket index)}; \label{fb_Z} \end{equation}
\vspace{-10pt}
\begin{equation} X^j \ket{+_k} = w^{jk} \ket{+_k}. \label{fb_X} \end{equation}

Let $\ket{\psi} = \sum_k a_k \ket{k}$. Then
\begin{eqnarray*}
\ket{\psi}\ket{+} &=& \sum_k a_k \ket{k} \ket{+} \\
& \stackrel{CZ}{\rightarrow} & \sum_k a_k \ket{k} \ket{+_k} \\
& \stackrel{F \otimes I}{\rightarrow} & \sum_k a_k \ket{+_k} \ket{+_k} \equiv \ket{\Psi}
\end{eqnarray*}
where we calculate the effect of the $CZ$ gate using equation (\ref{fb_Z}). Now, we can write $\ket{\Psi}$ as
\begin{eqnarray*}
\ket{\Psi} &=& \frac{1}{\sqrt{d}} \sum_{j,k} a_k \omega^{jk} \ket{j} \ket{+_k} \\
&=& \frac{1}{\sqrt{d}} \sum_j \ket{j} \sum_k a_k \omega^{jk} \ket{+_k} \\
&=& \frac{1}{\sqrt{d}} \sum_j \ket{j} \sum_k a_k X^j F\ket{k} \\
&=& \frac{1}{\sqrt{d}} \sum_j \ket{j} X^jF\ket{\psi} 
\end{eqnarray*}
where the penultimate line follows from equation (\ref{fb_X}) above; hence, when measuring the first qudit, an outcome $j$ (occuring with probability $1/d$) yields a state $X^jF\ket{\psi}$ for the second qudit.}

\subsection{Application to cluster state quantum computation}

One way we can think about the above circuit identity is that it gives us a very simple scheme for implementing the quantum Fourier gate $F$: Given a qudit pair in states $\ket{\psi}, \ket{+}$ entangled with a controlled-$Z$ interaction, if we measure the first ($\ket{\psi}$) qudit in the $F^\dagger$ basis, and obtain the measurement outcome $j$, the second qudit ends up in the state $X^j F \ket{\psi}$. The $X^j$ operator can be thought of as an \emph{error operator}, and we will discuss this shortly.

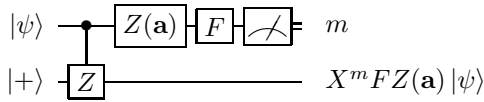
\begin{figure}[!htp] 
\[ 
\Qcircuit @C=0.4em @R=0.5em @!R{
\lstick{\ket{\psi}} & \ctrl{1} & \gate{Z(\mathbf{a})} & \gate{F} & \meter & \cw & \rstick{m} \\
\lstick{\ket{+}} & \gate{Z} & \qw & \qw & \qw & \qw & \rstick{X^m FZ(\mathbf{a}) \ket{\psi}} } 
\]
\caption{The more general circuit diagram for one-dit teleportation. \label{t2}}
\end{figure}

Furthermore, from equation (\ref{CZ}) we note that the controlled-$Z$ gate is symmetric in the two qudits, and so any \emph{phase} transformation in the computational basis (that is, a quantum gate of the form 
\[ Z(\mathbf{a})=\sum_k e^{ia_k}\oper{k}{k}, \] 
with $\mathbf{a} \in [0, 2 \pi]^d $) that acts on the first (upper) qudit commutes through the controlled-$Z$ gate. This means that measuring the first system in the basis defined by $(FZ(\mathbf{a}))^\dagger$, which (by lemma \ref{M}) is the same as applying $FZ(\mathbf{a})$ to the system before measurement in the computational basis, is equivalent to teleporting the initial state $Z(\mathbf{a})\ket{\psi}$. This leads to the more general circuit identity given in Figure \ref{t2}, meaning we can implement the quantum gates $FZ(\mathbf{a})$ using this method. 

If the transformation $\ket{\psi} \to FZ(\mathbf{a}) \ket{\psi}$ forms our entire quantum computation, then since the powers of $X$ merely permute the computational basis elements, and all known quantum algorithms conclude with a measurement in this basis, we can simply correct for the Pauli error $X^j$ via a \emph{classical} compuation. 

This is the simplest possible example of cluster state computation: We take two qudits, one of which is our \emph{initial} state $\ket{\psi}$, and an \emph{output} qudit, which is initialised in the state $\ket{+}$. After entangling the pair using a controlled-$Z$ interaction, and an appropriate measurement, we can map our input state $\ket{\psi}$ onto an output state $FZ(\mathbf{a})\ket{\psi}$, up to a Pauli error $X^j$. 

\begin{figure}[!bp]
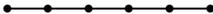
 
\[
\setlength{\unitlength}{0.3mm}
\centering
\bp(100,15)

\put(0,0){\line(1,0){90}}
\multiput(0,0)(18,0){6}{\circle*{3}}
\ep
\]
\caption{A linear cluster of 6 qudits. Each dot represents a physical qudit, and the lines an interaction between the two adjoined qudits. \label{c2}}
\end{figure}

\begin{figure}[!htp]

\[ 
\Qcircuit @C=0.4em @R=0.5em @!R{
\lstick{\ket{\psi}} & \ctrl{1} & \gate{Z(\mathbf{a})} & \gate{F} & \meter & \cw & \rstick{m_1} \\
\lstick{\ket{+}} & \gate{Z} & \qw & \qw & \qw & \qw & \qw & \ctrl{1} & \gate{Z(\mathbf{b})} & \gate{F} & \meter & \cw & \rstick{m_2} \\
 & & & & & & \lstick{\ket{+}} & \gate{Z} & \qw & \qw & \qw & \qw & \rstick{\ket{\psi^\prime}}  
 }
\]

\caption{A quantum circuit for two successive one-dit teleportations. The circuit diagram is arranged to illustrate that the teleportation between the upper two qudits is independent of any further structure in the cluster (in this case a third qudit entangled to the second). The final state will take the form $\ket{\psi^\prime} = X^{a(m_1,m_2)}Z^{b(m_1,m_2)}FZ(\mathbf{c}(\mathbf{b},m_1))FZ(\mathbf{a})\ket{\psi}$, where $\mathbf{c}$ is a function of $m_1$ because of the generalised Pauli error introduced by the first measurement may require the second measurement to be adapted. \label{1dt_many} }

\end{figure}
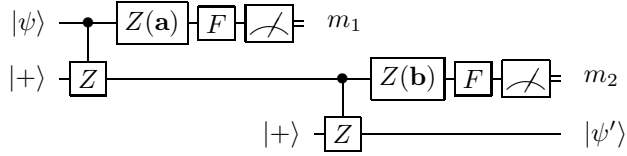

If we wished to apply further quantum operations to our state, we could do so by using a \emph{linear cluster} of qudits entangled in this way (Figure \ref{c2}), and then measuring along the cluster appropriately to implement the appropriate product of gates. Since measurement on a qudit and an interaction between another pair of qudits commute, this is equivalent to several one-dit teleportations, with the output from one teleportation becoming the input to the next (see Figure \ref{1dt_many}). The measurements after the first however may have to be adapted due to the extra $X$ factors brought in by previous measurements, so that the desired quantum operation is implemented. This issue of compensating for the randomness of previous measurement outcomes by a change of subsequent measurement bases is known as \emph{adaptive measurement}, and will be discussed in section 3. For now however, we will assume that this issue is surmountable. 

These ideas lead us to our definition of the cluster states we are going to use for computation:
\begin{enumerate}
\item We prepare a set of qudits, each in the state $\ket{+}$ (except perhaps for an `initial' set of qudits which are in an potentially entangled state $\ket{\psi}$; this represents the input to the algorithm);\footnote{Alternatively, some initial measurements on a cluster state where before any interactions are applied all qudits are initially in the state $\ket{+}$ could be used on a larger cluster to prepare this initial state.}
\item Finish creating the cluster by interacting 'neighbouring' pairs of qudits via a controlled-$Z$ gate.
\end{enumerate}
For qubits, this is identical to the original framework for cluster state QC \cite{CS1, CS2}. Cluster state computation then takes the form of a series of measurements on specified qudits, along with a scheme on how to adapt the measurements based on the previous measurement results. This definition also gives a possible method for generating these cluster states: suppose we have $d$-dimensional physical systems that we can prepare in the state $\ket{\psi}$, and we can allow them to interact with an appropriate interaction Hamiltonian between the systems e.g. 
\[ H_I = - g \sum_{k,l} kl \oper{k}{k} \otimes \oper{l}{l} \]
where $g$ is some coupling constant. If we let this interaction run for a time $t=2\pi/gd\hbar$, then the time evolution operator for the two qudits takes the form of a controlled-$Z$ gate. Physical systems which could be used to implement this scheme are discussed in \cite{dCQC} and appropriate references therein. For the remainder of this paper, we are going to discuss the mathematical framework of this model.

Before we continue, we make one final remark about our cluster states: they satisfy a property known as \emph{maximal connectedness}, which for our purposes we can state as follows: Given an arbitrary cluster state, if we measure any qudit of the cluster in the computational ($Z$) basis, the remaining states are left in a cluster state, up to some Pauli $Z$ errors. For qubit clusters, this issue is first discussed in \cite{CS_MC}. This fact can be used to design specific clusters from larger base clusters (e.g. a grid) as by the above, $Z$-measurements on a qudit effectively remove it from the cluster state. This is discussed in more detail in appendix \ref{app_maxc}.

\begin{figure}[!htp]
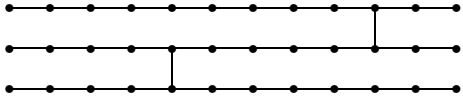
 
\[
\setlength{\unitlength}{0.3mm}
\centering
\bp(200,40)

\put(0,0){\line(1,0){198}}
\multiput(0,0)(18,0){12}{\circle*{3}}
\put(0,18){\line(1,0){198}}
\multiput(0,18)(18,0){12}{\circle*{3}}
\put(0,36){\line(1,0){198}}
\multiput(0,36)(18,0){12}{\circle*{3}}
\put(72,0){\line(0,1){18}}
\put(162,18){\line(0,1){18}}

\ep
\]
\caption{An example of a topology for cluster states that can be used to establish universal quantum computation. The rows represent a logical qudit, with one-dit teleportation the mechanism for implementing single-qudit gates. The interactions between rows are used to implement two-qudit gates, and so we can implement any arbitrary multi-qudit unitary operation. \label{c4}}
\end{figure}

Suppose we can use linear clusters to implement any single qudit operation. We can then use more general multi-dimensional clusters to allow the implementation of more complicated multi-qudit algorithms. One possible model is where rows of linear clusters, each representing a \emph{logical} qudit (i.e. corresponding to a single 'qudit wire' in the circuit model) are used to implement single qudit operations, and entanglement (again in the form of a controlled-$Z$ interaction) between the rows acts as an interaction between logical qudits. We will use this topology for cluster states a number of times throughout this paper; however, there are also a number of other ways of establishing this interaction between logical qudits.\footnote{The computational power of cluster states of many shapes are studied in \cite{CS_shape}.} It is a known fact that the set of one qudit gates along with any interacting two-qudit gate is sufficient to create any unitary gate on an arbitrary number of qudits (i.e. this set of gates is a \emph{universal} set for multi-qudit quantum gates) \cite{QCU}. Hence by establishing \emph{single}-qudit universality, we obtain universality for an arbitrary number of qubits on an appropriate cluster. A diagram to illustrate this point is given in Figure \ref{c4}.

We note that although in this paper we study single-qudit universality on linear clusters, since ultimately these ideas are to be applied to larger multi-dimensional clusters, we are not contradicting the theorem of Nielsen \cite{LCS} that says that quantum computation on linear clusters can be efficiently simulated on a classical computer (although the proof is for qubits, the same proof applies to qudits), as we ultimately want to use multi-dimensional clusters for multi-qudit QC.

For qubits, the set of gates given by $\mathcal{C}_2 = \{ FZ(\mathbf{a}) \ | \ \mathbf{a} \in [0, 2 \pi]^2 \}$ (i.e. those that can be implemented via one-\emph{bit} teleportation) is a \emph{universal} set for single-qudit quantum gates.\footnote{This is equivalent to the well known Euler decomposition for 2-dimensional unitaries.} In section \ref{sec_ung} we will consider the same question for the set $\mathcal{C}_d = \{ FZ(\mathbf{a}) \ | \ \mathbf{a} \in [0, 2 \pi]^d \}$.

Finally, we note that our definition of qudit cluster states is the same as that used in \cite{dCQC}. In that paper the stabiliser formalism for cluster state computation (introduced in \cite{CS2} and discussed later in section \ref{sec_stab}) is heavily used. We will use a much more direct approach to establish the theory.

\section{Adaptive measurements and adaptive computation} \label{sec_ad}

In this section we outline how the unwanted $X^m$ (generalised Pauli) operators that appear in our cluster state model can be corrected using classical computation, and by adapting, based on previous measurement results, the bases further measurements are taken in. This idea is known as \emph{adaptive measurements}. We will also show that by changing our correction method for these errors, we can extend one-dit teleportation to a further class of quantum gates; we call this \emph{adaptive computation}.

\subsection{Adaptive measurements for single qudit operations}

In using linear clusters and successive one-dit teleportation defined in the previous section to implement one-qudit quantum gates, generalised Pauli errors of the form $X^m$ are introduced to the overall state after each measurement. What we aim to show here is a remarkable feature of the original cluster state model that spreads to our generalisation, that we can compensate for these errors using solely \emph{classical} computation. 

Suppose we have a state $\ket{\psi} = X^{x}Z^{z}U\ket{\phi}$. The $U$ represents the quantum gate applied so far, and the powers of $X$ and $Z$ are the generalised Pauli errors that we have due to previous measurements. Suppose now we apply one-dit teleportation to the pair $\ket{\psi}\ket{+}$ entangled by a controlled-$Z$ operation as usual, and we obtain the new state $X^m F Z(\mathbf{a}) \ket{\psi}$. The final state can be written as $\ket{\psi^\prime}  =  X^m F Z(\mathbf{a}) X^{x}Z^{z}U\ket{\phi}$.The problem that we appear to have is that we are implementing the gate $F Z(\mathbf{a}) X^{x}Z^{z} U $ rather than the intended $F Z(\mathbf{a})U$. What we want is to move the powers of $X$ and $Z$ through the $F$ and $Z(\mathbf{a})$ so that they appear on the left of the overall gate $FZ(\mathbf{a})U$, so we can treat the Pauli operator as an error operator. To do this, we can make use of the following identities:
\begin{eqnarray}
Z(\mathbf{a}) X = X Z(\mathbf{a}^\prime); &\quad& (a^\prime_k = a_{k-1}) \label{Za_comm_X} \\
Z(\mathbf{a}) Z &=& Z Z(\mathbf{a}) \label{Za_comm_Z} \\
FZ =XF, &\quad& FX = Z^{-1}F \label{XZ} 
\end{eqnarray}
and hence (ignoring any changes in the irrelevant overall phase),
\begin{eqnarray}
\ket{\phi^\prime} & = & X^m F X^{x}Z^{z} Z(\mathbf{a}^{(x)}) U \ket{\phi} \nonumber \\
&=& X^m Z^{-x} X^{z} F Z(\mathbf{a}^{(x)}) U \ket{\phi} \nonumber \\
& = & \omega^{xz} X^{m+z} Z^{-x} F Z(\mathbf{a}^{(x)}) U \ket{\phi} \label{1dt_u}
\end{eqnarray}
where $\mathbf{a}^{(l)}$ is defined by $a^{(l)} = a_{k-l}$, and the final step is true because $XZ=\omega ZX$. Here we have succeeded in moving the generalised Pauli operators to where we want them; the only trouble now is that the overall gate that we are implementing is not identical to the one we want. We can get round this by using an \emph{adaptive} measurement: Given $x$ in advance, we can use one-dit teleportation to implement the gate $FZ(\mathbf{a}^{(-x)})$ instead of $F Z(\mathbf{a})$. With this adaptation, our new state is then given by (up to phase)
\[ \ket{\phi^\prime} = X^{x^\prime} Z^{z^\prime} F Z(\mathbf{a}) U \ket{\phi} \]
where
\begin{equation} x^\prime \equiv m + z \mod d; \quad
 z^\prime = -x \mod d.\end{equation}
So, by keeping track of the exponents of the $X$ and $Z$ operators by classical computation, and adapting the measurement taken via the original exponent values, we can apply the gate $F Z(\mathbf{a})$ to our state, up to a generalised Pauli error. The final step of the algorithm, which is (at least in all quantum algorithms to date) a measurement on the computational basis, can simply be performed without any quantum corrections to the final state, because the $Z$ errors have no effect on the probabilities of each outcome (they only change the computational basis vectors by a phase) and the $X$ errors permute the computational basis elements, which can be compensated for by a final classical computation.

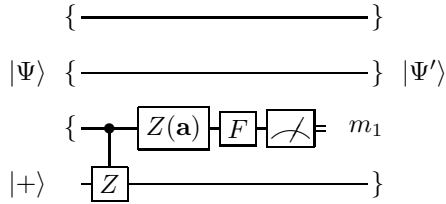
\begin{figure}[!bp]
\[  \Qcircuit @C=0.4em @R=0.5em @!R{
& \{ & & \qw & \qw & \qw & \qw & \qw & \qw & \qw & \qw & \qw & \}\\
\lstick{\ket{\Psi}} & \{ & & \qw & \qw & \qw & \qw & \qw & \qw & \qw & \qw & \qw & \} & \rstick{\ket{\Psi^\prime}} \\
& \{ & & \ctrl{1} & \gate{Z(\mathbf{a})} & \gate{F} & \meter & \cw & \rstick{m_1} \\
\lstick{\ket{+}} & & & \gate{Z} & \qw & \qw & \qw & \qw  & \qw & \qw & \qw & \qw & \} \\
}
 \]
\caption{One-dit teleportation applied to one qudit of a three-qudit state. It is easily seen (e.g. by writing $\ket{\Psi} = \Sigma_{ij} \ket{i}\ket{j} \ket{\psi_{ij}}$) that one-dit teleportation applied to one qudit of the multi-qudit state gives us the appropriate transformation on the transported qudit i.e. $\ket{\Psi^\prime} = (I \otimes I \otimes X^m FZ(\mathbf{a}))\ket{\Psi}$. \label{1dt_mq}}
\end{figure}

The above process would also work if we had a multi-qudit state $\ket{\Psi}$, with generalised Pauli errors acting on each qudit, and we entangled one of its qudits with a further $\ket{+}$ qudit via a controlled-$Z$ interaction, and then performed one-dit teleportation. This would simply implement a one-qudit operation on the given qudit, and change the generalised Pauli errors on that qudit also. This situation often arises when using more complicated multi-dimensional clusters, and is illustrated by Figure \ref{1dt_mq}.

We also note that we can move generalised Pauli errors through controlled-$Z$ gates, because of the identities
\begin{eqnarray}
CZ (Z \otimes I) = (Z \otimes I) CZ ; &\ &  CZ (I \otimes Z) = (I \otimes Z) CZ; \nonumber \\
CZ (X \otimes I) = (X \otimes Z^{-1}) CZ; &\ & CZ (I \otimes X) = (Z^{-1} \otimes X) CZ  \label{CZ_comm} \end{eqnarray}
which are easily established directly. These identities are necessary since we will sometimes need to treat a controlled-$Z$ interaction as an actual two-qudit gate, rather than a means to allow the use of one-dit teleportation. An example of this (from Figure \ref{c4}) is the vertical interaction between rows (logical qudits): we use these to create two-qudit interacting gates. The above relations mean that if two logical qudits are in a state $(X^{x_1}Z^{z_1} \otimes X^{x_2}Z^{z_2}) \ket{\Psi}$, applying a controlled-$Z$ gate to this and commuting it past the error operators yields a state 
\begin{equation} (X^{x_1}Z^{z_1-x_2} \otimes X^{x_2}Z^{z_2-x_1}) (CZ)\ket{\Psi}. \label{CZ_err} \end{equation}

\subsection{Adapative computation}

Given that a quantum algorithm ends in a measurement in the computational basis, generalised Pauli $X$ or $Z$ errors are not the only error operators that can be compensated for by a \emph{classical} compuation. Any overall operator that only permutes computational basis elements (and possibly modifies them by a phase) is also correctable by a final classical computation. 

Let us develop this idea further. Suppose we wish to implement a single unitary gate $U$ via a series of measurements on some cluster state, but end up with $Z^k P_\rho U$, where $\rho \in S_d$ and 
\begin{equation} P_\rho = \sum_{l=0}^{d-1} \oper{\rho(l)}{l}. \end{equation}
We can then compensate for this at the end of a quantum algorithm via classical computation also.

With this in mind, we can ask ourselves if we can somehow adapt our scheme for computation to include correction for more general permutation errors of the form above. We can express this more formally as follows: suppose we are given a state $\ket{\phi}=Z^n P_\rho U \ket{\psi}$, and we wish to apply the gate $FZ(\mathbf{a}) $ to the state via one-dit teleportation. We wish to establish whether we can peform a one-dit teleportation to obtain the state $\ket{\phi^\prime}=X^m F Z(\widetilde{\mathbf{a}})  Z^n P_\rho U \ket{\psi}$, and then rewrite this in the form $\ket{\phi^\prime} = Z^{m^\prime}P_{\rho^\prime} F Z(\mathbf{a}) U \ket{\psi}$, with $\rho^\prime \in S_d$. 

A simple class of permutations for which this is possible can be defined as follows. Let $c$ is a unit in the ring $\mathbb{Z}_d$.\footnote{A unit in a ring is an element that has a multiplicative inverse within the ring. $\mathbb{Z}_d$ is equal to the quotient $\mathbb{Z} / (d\mathbb{Z})$, and can be thought of as the ring of modulo $d$ arithmetic. In $\mathbb{Z}_d$, the units are the integers coprime to $d$ (i.e. have no common factor with $d$). Hence, when $d$ is prime, all non-zero elements are units; for non-prime dimensions not all elements are units e.g. 2 in $\mathbb{Z}_4$. If $c$ is a unit, then $cj=ck$ if and only if $j=k$. This property establishes $S_c$ as a permutation operator.} It then follows that the operator
\[ S_c = \sum_{k=0}^{d-1} \oper{ck}{k} \]
is a permutation operator. Furthermore, since
\begin{eqnarray*}
S_{c^{-1}}F &=& \frac{1}{\sqrt{d}} \sum_k \oper{c^{-1}k}{k} \sum_l \omega^{ml} \oper{m}{l} \\
&=& \frac{1}{\sqrt{d}} \sum_{l,m} \omega^{lm} \oper{c^{-1}m}{l} \\
&=& \frac{1}{\sqrt{d}} \sum_{l,m} \omega^{clm} \oper{m}{l} \\
&=& \frac{1}{\sqrt{d}} \sum_{l} \oper{+_{cl}}{l} \equiv F_c 
\end{eqnarray*}
it may be possible to implement the gate $F_c$ rather than $F$ by using the identity $F = S_cS_{c^{-1}}F = S_cF_c$, and treating the remaining $S_c$ permutation as an error operator (in the same way we treat powers of $X$ and $Z$). We use the phrase \emph{adaptive computation} to refer to this idea of implementing a different quantm gate by changing the leading error permutation that we correct for classically.

Let us suppose we are given a state $\ket{\psi} = X^{x}Z^{z}U\ket{\phi}$, which is in a linear cluster, and we measure it to induce a one dit teleportation, so the state becomes (by equation (\ref{1dt_u})) $\ket{\psi^\prime} = X^{m+z} Z^{-x} F Z(\mathbf{a}^{(x)}) U \ket{\phi}$ (if the measurement outcome is $m$). Note that the powers of the $X$ and $Z$ operators have already been updated. By applying the identity $F = S_cF_c$ here, we obtain the state
\begin{equation} \ket{\psi^\prime} = X^{m+z} Z^{-x} S_c F_c Z(\mathbf{a}^{(x)}) U \ket{\phi}. \label{1dt_err} \end{equation}
From this form we can see that by considering the $S_c$ operator as an error, our computational state is equal to $F_c Z(\mathbf{a}^{(x)}) U \ket{\phi}$, modulo a permutation error operator of the form $X^{x}Z^{z}S_c$. To be able to maintain this form after later one-dit teleportations and further uses of the identity $F = S_{c^\prime}F_{c^\prime}$ ($c^\prime$ any other unit in $\mathbb{Z}_d$), we need appropriate relations between $S_c$ and the operators $F$, $Z(\mathbf{a})$ that appear through the measurement process; however, it can easily be verified that
\begin{equation}  FS_c = S_{c^{-1}}F; \quad Z(\mathbf{a})S_{c} = S_{c}Z(\mathbf{a}^\prime) \ (a^\prime_k = a_{ck}). \label{S_Cgen} \end{equation}
which allow us to move the $S_c$ operator to the left of the intended unitary gate (the change in $\mathbf{a}$ above will mean adaptive measurement is needed here), and furthermore since $S_c S_d = S_{cd}$, it follows that we are always able to obtain an error operator of the form $X^{x}Z^{z}S_c$, where $c=1,\ldots,d-1$ and a unit in $\mathbb{Z}_d$. 

Note that we are still using the \emph{same} quantum process (one-dit teleportation), but a \emph{different} classical correction procedure. This means that we can implement the gates $F^\dagger Z(\mathbf{a})$ by choosing the measurement bases appropriately.

We note that for qubits, since the only unit in $\mathbb{Z}_2$ is $1=-1$, this means that $S_{-1}=I$, and since $F^\dagger=F$, this effect does not exist for cluster computation using qubits.

With this success for single-qudit operations, however, comes a word of warning for multi-qudit operations, through the identities
\begin{equation} CZ(S_c \otimes I) = (S_c \otimes I)C[Z^c]; \quad CZ(I \otimes S_c) = (I \otimes S_c)C[Z^c] \label{S_CgenZ}  \end{equation}
and since in our model the interactions between qudits are fixed, these changes in the effective interaction between e.g. neighbouring linear clusters are an artefact of this framework. However, since these extra permutation operators are \emph{not} introduced by a measurement process but instead through our choice in the classical computation, we can in theory design quantum algorithms to get round this problem or even utilise it to allow us to implement different interactions between neighbouring logical qudits. For example, since $S_{-1}^2 = I$, applying $F^\dagger$ rather than $F$ twice in a linear cluster produces two cancelling $S_{-1}$ factors which we no longer need worry about.

The group of permutations generated by the shifts $X^j$ and the multiplication maps $S_c$ is of order $\leq d(d-1)$, which is in general less than $d!$, the order of the symmetric group $S_d$. This means that the above does not deal with the most general permutation error. One issue with trying to correct for general permutation errors is that we would have to find a commutation relation between $P_\rho$ and $F$ for general $\rho$, which seems hard to ascertain.

We could have chosen to place the $S_c$ operator in front of the Pauli operators; in this case, we would need the following commutation relations:
\begin{equation} S_c X = X^c S_{c}; \quad S_c Z = Z^{c^{-1}} S_c. \end{equation}
These relations mean that $S_c$ (where $c$ is a unit in $\mathbb{Z}_d$) is a member of the \emph{generalised local Clifford group} $C_d$, which we define by the normaliser of the generalised Pauli group $\mathcal{P}_d = \{ \omega^a X^b Z^c \ | \ a,b,c \in \mathbb{Z}_d \}$ i.e.
\[ C_d = \{ U \in U(d) \ | \ UPU^\dagger \in \mathcal{P}_d \ \forall \ P \in \mathcal{P}_d \}. \]
In this case, $S_c$ corresponds to $U$, and products of powers of $X$ and $Z$ correspond to $P$. By looking in this group, we may find more permutations that we can correct for. However, it can be shown that in \emph{prime dimensions} the operators $Z, X, F, S_c$ and $P$ defined by $P: \ket{j} \to \omega^{j(j+1)/2}\ket{j}$ are sufficient to generate $C_d$ (see appendix \ref{app_c}). Since our permutations are already built out of products of $X$ and $S_c$, we cannot obtain any more \emph{single qudit} permutations from the Clifford group. This strongly suggests (but does not prove) that we cannot correct for any of the other single qudit permutation operators within cluster computation. It is harder to specify the Clifford group more generally, and so it may be the case that in non-prime dimensions one can implement further quantum gates by introducing other types of single qudit permutation error operators. 

\subsubsection{Adaptive computation using multi-level permutations}

When working with multi-qudit quantum algorithms, we can similarly introduce \emph{multi-qudit} permutation operators. Given an $n$ qudit state $EU\ket{\Psi}$, where $E$ represents an error operator on the $n$ qudits, and $U \in U(d^n)$ the desired quantum evolution acting on an initial state $\ket{\Psi}$, we can write 
\[ EU\ket{\Psi} = EP.P^{-1}U\ket{\Psi} \]
i.e. we let $P^{-1}$ become part of the quantum evolution, and $P$ part of the error operator $E$. To allow us to maintain this form of error operator we need $P$ to have appropriate relations with $F$ and tensor products of $Z(\mathbf{a})$. For example, consider a two qudit permutation $P$. After moving $P$ left through $(Z(\mathbf{a}) \otimes Z(\mathbf{b}))$, the operator to the right of $P$ msut be in tensor product form, so the transformations can be implemented using single qudit measurements. Up to permutations on individual systems, the only non-trivial permutation that does this is the swap operator $V$, since $V(A\otimes B) = (B \otimes A)V$. This gives us a swap of two qudits for free at any time, and the above relation for $V$ can be used to both update leading error operators and to calculate how measurement patterns need to be changed due to the introduction of the swap operator. However, we again have an issue with interacting gates between logical qudits changing form (for example, through the identity $CZ_{1,2}V_{23} = V_{23}CZ_{1,3}$; the indices represent the two systems each of these unitaries operates on). As before however, it could be that this is a help rather than a hinderance if utilised correctly. 
The swap can be introduced even in the qubit model, and, more generally, we can introduce any system swapping operator in this manner for an arbitrary number of qudits.\footnote{However, at the end of a computation before the final computational basis measurement, any permutation can be introduced in the above manner and corrected for classically, since there is no need to propogate the intoduced operators through other operators.}  


\subsection{Example - a variation on Deutsch-Josza}

Here we give an example of using one-dit teleportations to calculate the effect of a cluster state computation, and how adaptive computation can be useful in implementing quantum algorithms on qudit cluster states.

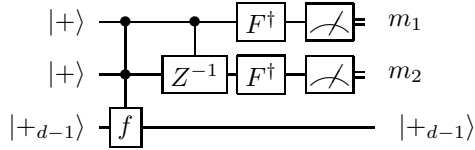
\begin{figure}[!htp] \[
\Qcircuit @C=0.4em @R=0.5em @!R{
\lstick{\ket{+}} & \ctrl{2} & \qw & \ctrl{1} & \gate{F^\dagger} & \qw & \meter & \cw & \rstick{m_1} \\
\lstick{\ket{+}} & \ctrl{1} & \qw & \gate{Z^{-1}} & \gate{F^\dagger} & \qw & \meter & \cw & \rstick{m_2} \\ 
\lstick{\ket{+_{d-1}}} & \gate{f} & \qw & \qw & \qw & \qw & \qw & \qw & \qw & \rstick{\ket{+_{d-1}}}  } \]
\caption{The quantum circuit for finding $a,b$. \label{DJ_qc}}
\end{figure}

Suppose we have a function $f: \mathbb{Z}_d^2 \to \mathbb{Z}_d; (x,y) \mapsto (x-a)(y-b)$, and we are interested in finding $a$ and $b$. Classically, this will require at least two evaluations of $f$. In this section we will show that the quantum algorithm presented in Figure \ref{DJ_qc} can find $(a,b)$ with one use of the function $f$, and that it can be implemented on a cluster state of few qudits.

Let $U_f$ be the unitary gate corresponding to the controlled evaluation of $f$ (the first gate in Figure \ref{DJ_qc}). It can easily be verified that
\begin{eqnarray*}
U_f \ket{x_1}\ket{x_2}\ket{+_{d-1}} &=& \omega^{f(x_1,x_2)}\ket{x_1}\ket{x_2}\ket{+_{d-1}} \\
&=& \omega^{ab}(Z^{-a}\otimes Z^{-b})CZ\ket{x_1}\ket{x_2}\ket{+_{d-1}} 
\end{eqnarray*}
and using this result it is easy to verify that this quantum algorithm maps the first two qudits to $\ket{-a}\ket{-b}$ (up to a phase), and so the final measurements yield the values of $(a,b)$.

\begin{figure}[!bp]
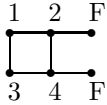
 
\[
\setlength{\unitlength}{0.3mm}
\centering
\bp(40,25)
\put(0,0){\line(1,0){36}}
\put(0,18){\line(1,0){36}}
\put(0,0){\line(0,1){18}}
\put(18,0){\line(0,1){18}}
\multiput(0,0)(18,0){3}{\circle*{3}}
\multiput(0,18)(18,0){3}{\circle*{3}}
\put(-1,-12){\textrm{3}}
\put(17,-12){\textrm{4}}
\put(35,-12){\textrm{F}}
\put(-1,23){\textrm{1}}
\put(17,23){\textrm{2}}
\put(35,23){\textrm{F}}
\ep
\]
\caption{The cluster state used to implement our quantum algorithm, complete with numbering for qudits (see text). The qudits labelled F represent the final two qudits at the end of the computation. \label{DJ_cs}}
\end{figure}

The form of the transformation induced by $U_f$ on the first two qudits is ideal for implementation on a cluster state depicted in figure \ref{DJ_cs}. The two rows will form the upper two (logical) qudits, and the vertical interactions will form interacting gates. In place of the unitary evaluation of $f$ will be a measurement on qudits 1 and 3, since we can write $Z^k = Z(\mathbf{z}^k)$ with $(z^k)_l = 2 \pi kl/d$. We will hence measure these qudits in the basis defined by $(FZ(\mathbf{z}^k))^\dagger$. To implement the $F^\dagger$ gates, we then measure in the $F^\dagger$ basis on qudits $2,4$, and use adaptive computation in the form $F=S_{-1}F_{-1} = S_{-1}F^\dagger$. Let $m_k$ be the result of the measurement on qudit $k$. We can then calculate the evolution of the cluster state algorithm (ignoring any overall phases), updating the error operators as we proceed using the relations established in the previous section:

\begin{eqnarray*}
\ket{+}\ket{+} & \stackrel{CZ}{\rightarrow} & CZ\ket{+}\ket{+} = \frac{1}{\sqrt{d}}\sum_k \ket{+}\ket{+_k} \\
& \stackrel{\textrm{measure 1,3}}{\rightarrow} & (X^{m_1} \otimes X^{m_3})(FZ(\mathbf{z}^{-a}) \otimes FZ(\mathbf{z}^{-b}))CZ\ket{+}\ket{+}  \\
& = & (X^{-a+m_1} \otimes X^{-b+m_3}S_{-1})(F \otimes F^\dagger)\frac{1}{\sqrt{d}}\sum_k \ket{+}\ket{+_k} \\
& = & (X^{-a+m_1} \otimes X^{-b+m_3}S_{-1})CZ\ket{+}\ket{+} \\
& \stackrel{CZ}{\rightarrow} & CZ(X^{-a+m_1} \otimes X^{-b+m_3}S_{-1})CZ\ket{+}\ket{+} \\
& = & (X^{-a+m_1}Z^{b-m_3} \otimes X^{-b+m_3}Z^{a-m_1}S_{-1})C[Z^{-1}]CZ\ket{+}\ket{+} \\
& \stackrel{\textrm{measure 2,4}}{\rightarrow} & (X^{m_2}F \otimes X^{m_4}F)(X^{-a+m_1}Z^{b-m_3} \otimes X^{-b+m_3}Z^{a-m_1}S_{-1})\ket{+}\ket{+} \\
& = & (X^{b+m_2-m_3}Z^{a-m_1}S_{-1} \otimes X^{a+m_4-m_1}Z^{b-m_3})(F^\dagger \otimes F^\dagger)\ket{+}\ket{+} \\
& = & \ket{-b-m_2+m_3 \mod d}\ket{-a-m_4+m_1 \mod d}  
\end{eqnarray*}

The values of $a$ and $b$ can then be read off from the final qudits. We note two things about the above computation:
\begin{enumerate}
\item The measurement of qudits 1 and 3 introduce extra $F$ gates on both logical qudits not present in the original algorithm. However, because $(F \otimes F^\dagger)CZ\ket{+}\ket{+} = CZ\ket{+}\ket{+}$, by using the identity $F=S_{-1}F^\dagger$ on one qudit, these extra gates do not affect the rest of the computation, except for the fact that the first qudit corresponds to the value of $b$ rather than $a$ as in the circuit of Figure \ref{DJ_qc} (and vice versa).
\item The original algorithm requires the use of a controlled-$Z^{-1}$ gate, where as our cluster state only has controlled-$Z$ gate between the logical qudits. However, pulling through the $S_{-1}$ error operator (introduced in the previous step) to the left through the $CZ$ gate between qudits 2 and 4, the $CZ$ gate changes into a $C[Z^{-1}]$ gate.
\end{enumerate}

While this example thought of purely as a quantum algorithm is a bit of a toy example, it does illustrates not only how adaptive computation can be helpful in the implementation of algorithms on cluster states, but also how the non-trivial commutation relation between $S_c$ and $CZ$ can actually prove to be of help rather than a hinderance.

\section{The stabiliser formalism for cluster state QC} \label{sec_stab}

Up to this point, we have presented a completely self-contained approach to cluster state computation. Most of the current papers \cite{CS1, CS2, dCQC} take a much more algebraic approach using the \emph{stabiliser formalism}, developed by Gottesman \cite{SF} in the context of error-correcting quantum codes. The stabiliser formalism allows us to describe cluster states through sets of eigenvalue equations, and manipulation of these eigenvalue equations leads to a theorem that can be used to establish how a particular set of measurements can implement a quantum gate on a particular cluster state. In this section we will briefly summarise this approach, and compare the merits of the two.

\subsection{Representation of cluster states using stabilisers}

Before we describe how stabilisers can be used to represent cluster states, let us give a more formal definition of cluster states. Let $G=(V,E)$ be a graph; the set $V$ are the vertices (this corresponds to a labelling of our physical qudits) and $E \subseteq V \times V$ are the edges of $G$ (corresponding to which pairs of vertices have a controlled-$Z$ interaction between them). For $a \in V$, let $\ket{.}_a$ represent the state of qudit $a$. Then our cluster state $\ket{\phi}_{C(G)}$ can be written as
\[ \ket{\phi}_{C(G)} = \left( \prod_{(a,b) \in E} CZ_{(a,b)} \right) \left( \bigotimes_{a \in V} \ket{+}_a \right) \]
where $CZ_{(a,b)}$ represents the controlled-$Z$ gate between qudits $a$ and $b$. 

It can be shown \cite{CS2, dCQC} that $\ket{\phi}_{C(G)}$ is uniquely determined (up to phase) by the eigenvalue equations
\[ X^\dagger_a \otimes \left( \bigotimes_{b \in N(a)} Z_b \right) \ket{\psi} = \ket{\psi} \]
where $U_a$ is the application of the unitary $U$ to qudit $a$, and $N(a)= \{ b \in V \ | \ (a,b) \in E \}$ represents the neighbours of qudit $a$ i.e. the qudits $b$ connected to $a$ by an interaction (edge in G). The operators $S(a) = X^\dagger_a \otimes \left( \bigotimes_{b \in N(a)} Z_b \right)$ are the \emph{stabilisers} of the state $\ket{\phi}_{C(G)}$, and totally determine up to an irrelevant phase the cluster state.

\subsection{Measurement patterns in the stabiliser formalism}

Before we can state the main result that shows how stabilisers can be used to help find measurement patterns that implement useful quantum algorithms, we need to establish a few definitions. Given a cluster state $\ket{\phi}_{C(G)}$, let us break its graph up into three parts: an input cluster $C_I(G)$ and output cluster $C_O(G)$, and the remaining body of the cluster, which we will denote $C_M(G)$. We will number the $n$ qudits in $C_I(G)$ and $C_O(G)$ from $1$ to $n$, where qudit $i$ in the output cluster can be thought of as qudit $i$ from the input cluster after the quantum computation has concluded.  We suppose further that the $n$ qudits in $C_I(G)$ can, before any interactions between qudits is applied, be initialised in the more general entangled state $\ket{\psi(\textrm{in})}$. 

We define a \emph{measurement pattern} on a cluster $C(G)$ to be a set of the form of a function $M_{C(G)} : V \to U(d)$, such that the columns of the unitary matrix $M_{C(G)}(a)$ form the measurement basis on qudit $a$ in the cluster.\footnote{This definition differs slighlty from that used in other papers but is completely equivalent.} If the measurement outcomes form a vector $\mathbf{m}$, let $P(M_{C(G)}, \mathbf{m})$ represent the projector onto the states corresponding to the particular measurement outcomes.

We are now ready to state the main theorem for cluster state QC and stabilisers:
\vspace{10pt}
\begin{thm}[\cite{CS2,dCQC}] Suppose the state $\ket{\psi}_{C(G)} = (I_{C_I(G)} \otimes P(\mathcal{M}_{C(G)}, \mathbf{m}) \otimes I_{C_O(G)})\ket{\phi}_{C(G)}$ obeys the $2n$ eigenvalue equations
\begin{eqnarray}
X_{i,C_I(G)} (UX_iU^\dagger)_{C_O(G)} \ket{\psi}_{C(G)} &=& \omega^{\lambda_{x,i}} \ket{\psi}_{C(G)} \\
Z^\dagger_{i,C_I(G)} (UZ_iU^\dagger)_{C_O(G)} \ket{\psi}_{C(G)} &=& \omega^{\lambda_{z,i}} \ket{\psi}_{C(G)}.
\end{eqnarray}
Then, if $\mathcal{M}_{C_I(G)} = \{(i,F_i) \ | \ i \in C_I(G) \}$, and the measurement outcomes on $C_I(G)$ are given by $s_i$, then after all measurements on $C_I(G)$ and $C_M(G)$ of $\ket{\phi}_{C(G)}$, the final state $\ket{\psi(\textrm{out})}_{C_O(G)}$ is given by $\ket{\psi(\textrm{out})}_{C_O(G)} = U U_e \ket{\psi(\textrm{in})}$, where 
\[ U_e = \bigotimes_{i=1}^n (Z_i)^{\lambda_{x,i}-s_i}(X_i)^{\lambda_{z,i}}. \]
\end{thm}
\vspace{10pt}
The unitary gate $U$ and the numbers $\lambda_{x,i}, \lambda_{z,i}$ can all depend on the outcomes of the measurements on the input and body parts of the cluster state. Many examples of the use of this formalism can be found in \cite{CS2, dCQC}. 

\subsection{A comparison of the two approaches to cluster state QC}

The two approaches to cluster state QC that we have mentioned are very different. The stabiliser formalism is a powerful tool in establishing cluster states and measurements on them to implement particular quantum gates, but as examples in \cite{CS2} show, using it is an involved and difficult process. The one-dit teleportation approach described here is a much more elementary description of the model, and it has two important advantages over the stabiliser approach. Since this approach is very closely linked to the circuit model for quantum computation, it means that it is possible to use some of the intuition and results from this line of thought in designing quantum algorithms using cluster states. This approach is also the easiest to understand as it uses no more than some basic linear algebra to establish the theory behind the model.

\begin{figure}[!tp]
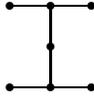
 
\[
\setlength{\unitlength}{0.3mm}
\centering
\bp(40,40)

\put(0,0){\line(1,0){36}}
\put(0,36){\line(1,0){36}}
\put(18,0){\line(0,1){36}}

\multiput(0,0)(18,0){3}{\circle*{3}}
\multiput(0,36)(18,0){3}{\circle*{3}}
\put(18,18){\circle*{3}}
\ep
\]
\caption{An example of a cluster that we cannot evaluate using one-dit teleportations alone. \label{1dt_bad}}
\end{figure}

The fact however that the one-dit teleportation approach is so closely linked to the ciruit model means that this approach is not suitable for considering all possible cluster states. Up to this point, we have always implicitly considered clusters that are in a grid (two-dimensional), such that each row of entangled qudits corresponds to a logical qudit, to which we apply successive one-dit teleportations, and any vertical interactions (columns) form two-qudit gates between logical qudits (as in Figure \ref{1dt_many}). However, clusters where a row does not form a logical qubit of the computation, but forms part of an interaction between logical qudits, cannot have one-dit teleportation ideas applied to it. The cluster in Figure \ref{1dt_bad} is an example of this. The central qudit causes the problem: its presence means that there are not two disjoint paths from left to right through the cluster such that every qudit is in one path, so we cannot treat the cluster by implementing one-dit teleportations along the paths (to give single qudit operations) and considering links between the paths as interacting gates.

Another issue with the one-dit teleportation approach is that of the timings of the measurements. In all of our clusters so far, all measurements in a row of entangled qudits must be taken `from left to right' i.e. we must apply each teleportation in turn. While we could make joint measurements of qudits where no adaptive measurements take place, we cannot measure the qudits out of turn in the teleportation approach. Examples in \cite{CS2} using the stabiliser method however distinctly allow measurements to be taken in an arbitrary order (up to the issue where some measurements may depend on other measurement results). 

Importantly though, the tool of adaptive computation does not naturally arise within the stabiliser formalism. It seems difficult to incorporate the extra operator errors that arise from adaptive computation into the major theorem above (particularly as their deterministic origin differs from the randomness of the origin of the Pauli operators), and it remains to be seen if there is an elegant or natural way that these two methods could be combined. The fact that the same quantum process is taking place but a different correction procedure is implemented would somehow have to be reflected in any such grander theorem.

\section{Universal gates} \label{sec_ung}

In this section we are going to discuss how we can implement any unitary evolution on a single qudit using linear clusters of qudits. Universality for many qudits using multi-dimensional clusters follows from the comments in section 2. 

\subsection{An approach to generating unitary gates}

In this section we are going to present a couple of basic results that present a possible method of generating unitary transformations on any QC system.

We start by quoting the following result:
\vspace{10pt}
\begin{lem}[p48, \cite{EXP}] Let $H_k$ be a basis over the reals for the space of $d \times d$ Hermitian matrices. Then any $U \in U(d)$ can be formed by products of the matrices $\exp (i \beta_k H_k)$ for real $\beta_k$. \label{Hsp} \end{lem}
\vspace{10pt}
Now, given a matrix $U$, with eigenvectors $\ket{U_j}$, we will define
\begin{equation} U(\mathbf{a}) = \sum_k e^{ia_k} \oper{U_k}{U_k}. \end{equation}
Suppose we can find a (minimal) set of matrices $U_1, \ldots, U_m$ such that the projectors onto the eigenvectors of these matrices form (over the reals) a spanning set for the space of $d \times d$ hermitian matrices. It follows from lemma (\ref{Hsp}) that products of the unitaries from the set
\[ \mathcal{C} = \{ U_j(\mathbf{a}) | j \in \{1,\ldots,d \}, \ \mathbf{a} \in [0, 2 \pi]^d  \} \]
is enough to generate any unitary in $U(d)$.\footnote{Note that we not using operators of the form e.g. $exp(i\alpha Z)$ for some real $\alpha$, as is used for qubits. This is because for $d \geq 3$, $Z, X$ are not hermitian, and so $exp(i\alpha Z),exp(i\alpha X)$ are not unitary matrices.}

This result in itself does not specifically apply to our model of QC yet; we need to add more constraints to the kinds of bases that we can use.

\subsection{Mutually unbiased bases and Hermitian matrices}

To motivate this discussion, we will state here (and prove later) that, using a linear cluster of three qudits and two measurements, with the first (left-most) qudit in a given state $\ket{\psi}$, we can implement the gates $Z(\mathbf{a})$ and $X(\mathbf{a})=\sum_k e^{ia_k} \oper{+_k}{+_k}$ for any dimension $d$. The two bases $\{ \ket{k} \}_{k=0}^{d-1} $ and $\{ \ket{+_k} \}_{k=0}^{d-1} $ are what we call \emph{mutually unbiased}, a property defined as such:
\vspace{0.1pt}
\begin{defn} Two bases $\{ \ket{a_k} \}_{k=0}^{d-1} $ and $\{ \ket{b_k} \}_{k=0}^{d-1}$ are \textbf{mutually unbiased} if 
$|\inprod{a_i}{b_j}| = 1/\sqrt{d}$ for all $i,j$. \end{defn}
\vspace{4pt}
Indeed, if we accept one-dit teleportation as the basis of the cluster state model, the only bases that we can measure are those that are mutually unbiased with respect to the standard basis. We can take this connection further with the following observation:
\vspace{4pt}
\begin{lem} Let $\{\ket{\psi_j^k}\}_{j=1}^d$ be $d+1$ bases of $\mathbb{C}^d$ (with $k=1,\ldots,d+1$), such that they are mutually unbiased. Then real combinations of projectors of all these vectors span the space of $d \times d$ Hermitian matrices. \end{lem}
\vspace{4pt}
This arises from the fact that mutually unbiased bases arise in the problem of quantum state determination. In the limit of many measurements, measurements in each of a full set of $d+1$ mutually unbiased bases are enough to specify any Hermitian matrix \cite{MUB}. So mutually unbiased bases are potentially a route into finding a basis of projectors for the set of Hermitian matrices that are appropriate to our model. 

We will list briefly some known facts about mutually unbiased bases. It is known \cite{MUB,MUB2} that the maximum number of possible mutally unbiased bases in dimension $d$ is $d+1$. In dimensions that are prime powers, a construction is known that gives this maximal possible number of such bases \cite{MUB,MUB2}. However, more generally no construction is known to find such a maximal set in arbitrary dimension. For the first non-prime power case, $d=6$, there are 3 known MUBs, but it is not known if any larger set exists. 

\subsection{An explicit construction for prime dimensions}

For dimension $d$ a prime, it can be shown via a result of \cite{MUB2} that a set of $d+1$ mutually unbiased bases can be obtained via the eigenvectors of the $d+1$ matrices
\[ Z, X, ZX, ZX^2, \ldots, ZX^{d-1}. \]
The bases $\{ \ket{k} \}_{k=0}^{d-1} $ and $\{ \ket{+_k} \}_{k=0}^{d-1} $ are the eigenvectors of $Z$ and $X$ respectively. 

What we will show in this section is that we can generate the gates $Z(\mathbf{a})$, $X(\mathbf{a})$ and $ZX^k(\mathbf{a})$ for prime dimensions on linear clusters. Then, by the results of the previous two sections, this will establish single-qudit universality on linear clusters as required.

It is easy to implement $Z(\mathbf{a})$ and $X(\mathbf{a})$ in any dimension using two measurements on a linear cluster of size three, because we may write 
\[ Z(\mathbf{a}) = F^\dagger . FZ(\mathbf{a}); \quad X(\mathbf{a}) = FZ(\mathbf{a}).F^\dagger. \]
and hence by appropriate one-dit teleportations and use of adaptive computation we can implement these gates.\footnote{Note that the gate $F^\dagger$ can be implemented without the use of adaptive computation with three one-dit teleportations, since $F^\dagger = F^3$, but this takes up more resources.} We note that this scheme uses fewer measurements to implement these gates than the one given in \cite{dCQC}. These two gates will be the building blocks for implementing the gates $ZX^k(\mathbf{a})$. The following lemma is crucial in allowing us to do this:
\vspace{6pt}
\begin{lem} \label{EP} Let $\{ \ket{\psi^k_j} \}_{j=1}^d $ be the eigenvectors of $ZX^k$ for $k=1, \ldots, p-1$, with eigenvalues $\omega^j$. There exists a phase transformation $Z(\mathbf{b}_k)$ such that
\[ Z(\mathbf{b}_k) \ket{+_j}  = \ket{\psi^k_{jk}}. \]
\end{lem}
\vspace{2pt}
The index $jk$ in the ket determines which eigenstate of $ZX^k$ the phase transformation $Z(\mathbf{b}_k)$ maps $\ket{+_j}$ onto; importantly, the index is \emph{not} solely $j$, and depends on $k$. The proof of the lemma is a little technical and hence we postpone it to appendix \ref{app_l}. The result of this is that we may write
\[ ZX^k(\mathbf{a}) = Z(\mathbf{b}_k) X(\mathbf{a}^{(\times,k)}) Z(\mathbf{b_k})^\dagger \]
where $\mathbf{a}^{(\times,k)}$ is a permutation of the elements of $\mathbf{a}$ defined by $a^{(\times,k)}_l = a_{k^{-1}l}$. Hence we can decompose $ZX^k(\mathbf{a})$ as
\[ ZX^k(\mathbf{a}) = F^\dagger. FZ(\mathbf{b}_k). FZ(\mathbf{a}^{(\times,k)}).F^\dagger Z(\mathbf{-b_k}) \]
and so we can implement these gates using four measurements and use of adaptive computation.

\subsection{The $d=3$ case}

In this section we will give the above results in dimension 3 in more detail, to show explicitly which measurements are needed to implement each gate.

In dimension 3, we have four primitive gates in our construction: $Z(\mathbf{a}), X(\mathbf{a}), ZX(\mathbf{a})$ and $ZX^2(\mathbf{a})$. We need the eigenvectors of each of the matrices $Z, X, ZX$ and $ZX^2$. The computational basis elements are the eigenvectors of $Z$. The eigenvectors of $X$ are the columns of $F$ i.e.
\[\begin{array}{ccc}
\ket{+_0} &=& \frac{1}{\sqrt{3}} \left(  \ket{0} + \ket{1} +  \ket{2} \right) \\
\ket{+_1} &=& \frac{1}{\sqrt{3}} \left(  \ket{0} + \omega\ket{1} +  \omega^2\ket{2} \right) \\
\ket{+_2} &=& \frac{1}{\sqrt{3}} \left(  \ket{0} + \omega^2\ket{1} +  \omega\ket{2} \right); 
\end{array} \]
the eigenvectors of $ZX$ are 
\[\begin{array}{ccc}
\ket{ZX_0} &=& \frac{1}{\sqrt{3}} \left(  \ket{0} + \ket{1} +  \omega^2\ket{2} \right) \\
\ket{ZX_1} &=& \frac{1}{\sqrt{3}} \left(  \omega^2\ket{0} + \ket{1} +  \ket{2} \right) \\
\ket{ZX_2} &=& \frac{1}{\sqrt{3}} \left(  \ket{0} + \omega^2\ket{1} +  \ket{2} \right); 
\end{array} \]
and the eigenvectors of $ZX^2$ are
\[\begin{array}{ccc}
\ket{ZX^2_0} &=& \frac{1}{\sqrt{3}} \left(  \ket{0} + \omega\ket{1} +  \ket{2} \right) \\
\ket{ZX^2_1} &=& \frac{1}{\sqrt{3}} \left(  \ket{0} + \ket{1} +  \omega\ket{2} \right) \\
\ket{ZX^2_2} &=& \frac{1}{\sqrt{3}} \left(  \omega\ket{0} + \ket{1} +  \ket{2} \right). 
\end{array} \]
By utilising the proof of lemma \ref{EP} in appendix \ref{app_l} (or indeed by direct verification) defining $\mathbf{b}_1 = (0,0,4\pi/3)^T$ and $\mathbf{b}_2 = (0, 2\pi/3, 0)^T$, so that
\[ Z(\mathbf{b}_1) = \left( \begin{array}{ccc} 1 & 0 & 0 \\ 0 & 1 & 0 \\ 0 & 0 & \omega^2 \end{array} \right) \quad Z(\mathbf{b}_2) = \left( \begin{array}{ccc} 1 & 0 & 0 \\ 0 & \omega & 0 \\ 0 & 0 & 1 \end{array} \right)
 \]
we have that
\begin{eqnarray*}
Z(\mathbf{b}_1)\ket{+_0} &=& \ket{ZX_0} \\
Z(\mathbf{b}_1)\ket{+_1} &=& \omega \ket{ZX_1} \\
Z(\mathbf{b}_1)\ket{+_2} &=& \ket{ZX_2} 
\end{eqnarray*}
and
\begin{eqnarray*}
Z(\mathbf{b}_2)\ket{+_0} &=& \ket{ZX^2_0} \\
Z(\mathbf{b}_2)\ket{+_1} &=& \omega^2 \ket{ZX^2_2} \\
Z(\mathbf{b}_2)\ket{+_2} &=& \ket{ZX^2_1}.
\end{eqnarray*}
From these relations we can deduce that
\[ ZX(\mathbf{a}) = Z(\mathbf{b}_1) X(\mathbf{a}) Z(\mathbf{b_1})^\dagger \]
and
\[ ZX^2(\mathbf{a}) = Z(\mathbf{b}_2) X(\mathbf{a}^\prime) Z(\mathbf{b_2})^\dagger \]
where $\mathbf{a}^\prime$ is defined by
\[ a^\prime_1 = a_1; \quad a^\prime_2 = a_3; \quad a^\prime_3 = a_2 \]
and using the measurement patterns in the previous subsection we can implement these gates on linear clusters.

The issue of adaptive computation in this dimension is simple, because only when $c=-1$ is $S_c$ unitary and not equal to the identity. So we can implement the gates $FZ(\mathbf{a})$ and $F^\dagger Z(\mathbf{a})$ in this dimension, and by doing so we only introduce one new kind of error, namely that introduced by $S_{-1}$.

Furthermore, it is easy to show that the shift permutation $(3 1 2)$ and the transposition $(2 3)$ generate the whole of the permutation group $S_3$; these permutations correspond to $X$ and $S_{-1}$ and so when $d=3$ we can in fact correct for all possible permutation errors.

\subsection{Other cases}

As noted above, constructions for full sets of mutually unbiased bases exist in all prime power dimensions, but their constructions are not as simple as the one listed above. In \cite{MUB2}, an explicit description for a set of mutually unbiased bases in $d=4$ is given, and can be seen by inspection that a relationship similar to that for prime dimensions given by lemma (\ref{EP}) can be given. It remains to be seen whether this relationship exists more generally.

Furthermore, these mutually unbiased bases should be compatible with our scheme for generating universal gates in the cluster state model. Ideally one of the bases should coincide with the Fourier basis (\ref{Fb}), to fit in with the $F$ transformation from one-dit teleportation. This is not the case in the example in \cite{MUB2}, which makes the task of generalising the ideas in this paper to non-prime dimensions more difficult.

It is clearly not the case however that non-existence of a full set of MUBs implies that our cluster state model is not universal. It is established in \cite{dCQC} that our given model is universal for all dimensions. They also prescribe a method which is equivalent to finding a set of projectors that is not only a basis for the set of Hermitian matrices, but also suitable for the cluster state model. It remains to be seen whether the model can be described more directly as in this paper.

\section{Conclusion}

In this paper we have developed a framework for quantum computation using cluster states with qudits as our basic physical resource, and while we have recovered all of the features of the qubit model, we have found that the extra degrees of freedom in using qudits allows us to control the quantum computation using adaptive computation in a way that is not possible (for single qudit operations) when using qubits. The cases of prime dimension lend themselves particularly well to a simple description, and as a result it is hopeful that these ideas could be taken further and used to design quantum algorithms on qudits.

One important issue we have not covered in this paper is that of fault tolerance. There are a number of works on this issue specifically pertaining to the qubit cluster state model \cite{CS_F, CS_F2, R_PhD}, and given the similarities between this and the qudit model, it seems likely that a number of the results may well jump across to the qudit model. How the issue of adaptive computation will fit into the picture could however be a sticking point for generalising these results.

The most technologically advanced physical implementation of cluster state QC is an optical solution as proposed by Nielsen \cite{CS_O} and refined by Browne and Rudolph \cite{CS_O2}. A photon can be thought of as a qubit by treating the polarisation of the photon as the relevant degrees of freedom, and in \cite{KLM} Knill, Laflamme and Milburn presented a scheme for entangling photons using measurement. This idea is at the center of Nielsen's proposal; Browne and Rudolph present another  optical scheme that is more efficient in its use of resources (linear optical elements).\footnote{Expermiental work using photonic cluster states has been reported in \cite{CS_E}.} However, by using other degrees of freedom present for photons, they could be potentially thought of as qudits. It would most definitely be interesting if one could find a scheme that is similar to any of the above proposals, but within which photons are treated as qudits, and furthermore, whether this idea lends itself to a scheme for qudit cluster state quantum computation.

A number of other interesting questions arise from this model. One immediate question is whether the computational paradigm that arises from adaptive computation lends itself to solving a particular class of problems. Furthermore, we have not explored the potential for designing clusters with controlled-$Z^k$ interactions, as was mentioned in section 3. Finally, we have not made any attempt here to consider fundamentally the role of entanglement within the cluster state model. This encompasses a large number of potential questions, and further investigation could lead to some insight into the relationship between computation and entanglement.

\section*{Acknowledgements}
The author would like to thank Tony Sudbery for reading this manuscript and his continual support throughout the research, Terry Rudolph whose presentation on cluster states encouraged the author to tackle this subject, Simone Severeni for a useful discussion, Sam Braunstein for the use of his library, Paul Butterley and Calvin Smith for pointing out a number of mistakes in the final draft, and the Engineering and Physical Science Research Council (U.K.) for financial support. The circuit diagrams were created using Qcircuit, available at \url{http://info.phys.unm.edu/Qcircuit}.

\appendix
\section{Maximal connectedness of cluster states} \label{app_maxc}

This appendix establishes the result that if any one qudit in a cluster is measured, the remaining qudits form a cluster state up to some Pauli $Z$ errors. As mentioned in the body of the text, for cluster of qubits this issue was first discussed in \cite{CS_MC}; here we give a different but elementary proof of the same fact. 

Let $\ket{\psi}$ be a cluster state. Since all the controlled-$Z$ interactions between pairs of qudits commute, we can write this as 
\[ \ket{\psi} = S \ket{+} \ket{\psi^\prime} \]
where $\ket{\psi^\prime}$ is the cluster state of all but one of the qudits, and $S$ represents the controlled-$Z$ interactions with the remaining $\ket{+}$ qudit. We can write the above as
\begin{eqnarray*}
\ket{\psi} &=& \frac{1}{\sqrt{d}} S \sum_{k=0}^{d-1} \ket{k} \ket{\psi^\prime} \\
					 &=& \frac{1}{\sqrt{d}} \sum_{k=0}^{d-1} \ket{k} (Z^k \otimes \ldots \otimes Z^k \otimes I \otimes \ldots \otimes I) \ket{\psi^\prime} 
\end{eqnarray*}
where the $Z^k$ operators act only on the qudits that are immediately connected to the final qudit (that is, those that have been made to interact with the final qudit by a controlled-$Z$ operation). It is clear from this form that measuring this qudit in the computational basis and obtaining result $j$ will project the rest of the state onto a cluster state with Pauli $Z^j$ errors on all qudits that were adjoined to the measured qudit in the original cluster.

Since the cluster state of two qudits is equivalent (up to unitaries on the individual qudits) to a Bell state, our notion of maximal connectedness is identical to that discussed for linear clusters in \cite{CS_MC}.

\section{Proof of lemma \ref{EP}} \label{app_l}

To prove the lemma, we first need to compute the eigenvectors of $ZX^k$ in prime dimensions $d \geq 3$.
\begin{lem} Define $\ket{\mathbf{\underline{\alpha}}^{(k)}}$ by 
\[ \ket{\mathbf{\underline{\alpha}}^{(k)}} = \frac{1}{\sqrt{d}} \sum_{l=0}^{d-1} \omega^{\alpha_l} \ket{l} \]
where $\underline{\alpha}$ satisfies $\alpha_{l+k} + l \equiv \alpha_l \mod d$ (with mod $d$ arithmetic in indices also). Then 
$\ket{\mathbf{\underline{\alpha}}^{(k)}}$ is an eigenvector of $ZX^k$ with eigenvalue 1.
\end{lem}
\proof{
Rewrite $\ket{\mathbf{\underline{\alpha}}^{(k)}}$ as
\[\ket{\mathbf{\underline{\alpha}}^{(k)}} = \frac{1}{\sqrt{d}} \sum_{l=0}^{d-1} \omega^{\alpha_{l+k}} \ket{l+k} \]
(with modulo $d$ addition as appropriate). Then
\begin{eqnarray*}
ZX^k\ket{\mathbf{\underline{\alpha}}^{(k)}} &=& \frac{1}{\sqrt{d}} \sum_{l=0}^{d-1} \omega^{\alpha_{l+k}+l} \ket{l} \\
&=& \frac{1}{\sqrt{d}} \sum_{l=0}^{d-1} \omega^{\alpha_l} \ket{l} \\
&=& \ket{\mathbf{\underline{\alpha}}^{(k)}}.
\end{eqnarray*} 
The other eigenvectors of $ZX^k$ are then given by $X^{-m}\ket{\mathbf{\underline{\alpha}}^{(k)}}$, since
\begin{eqnarray*}
ZX^k . X^{-m}\ket{\mathbf{\underline{\alpha}}^{(k)}} &=& \omega^{m} X^{-m}. ZX^k \ket{\mathbf{\underline{\alpha}}^{(k)}} \\
&=& \omega^{m} X^{-m} \ket{\mathbf{\underline{\alpha}}^{(k)}} 
\end{eqnarray*}
For definiteness, we define the phases of the eigenvectors by letting $\ket{\psi^k_0} = \ket{\mathbf{\underline{\alpha}}^{(k)}}$, with $\alpha^{(k)}_0=0$, and $\ket{\psi^k_j} = X^{-j}\ket{\mathbf{\underline{\alpha}}^{(k)}}$. The following lemma is equivalent to lemma \ref{EP}:
\vspace{10pt}
\begin{lem} Let $\{ \ket{\psi^k_j} \}_{j=0}^{d-1} $ be the eigenvectors of $ZX^k$ for $k=1, \ldots, p-1$ as defined above. Then there exists a phase transformation $Z(\mathbf{b}_k)$ such that
\[ Z(\mathbf{b}_k) \ket{+_j}  = \omega^{\frac{1}{2}j(j+1)k} \ket{\psi^k_{jk}}. \]
\end{lem}
\vspace{10pt}
\proof{ By the above lemma, 
\[ \ket{\psi^k_m} = \frac{1}{\sqrt{d}} \sum_{l=0}^{d-1} \omega^{\alpha_l} \ket{l+m} \]
where $\alpha_{l+k} + l \equiv \alpha_l \mod d$. Define $\mathbf{b}_k$ by $(b_k)_l = \alpha_l$. Then 
\[ Z(\mathbf{b}_k)\ket{+_{-j}} = \frac{1}{\sqrt{d}} \sum_{l=0}^{d-1} \omega^{\alpha_l - jl} \ket{l}. \]
However, it is easily seen (e.g. by induction) that, for any $r,s$, 
\begin{equation} \alpha_l \equiv \alpha_{l+jk} + jl + \frac{1}{2}j(j-1)k  \label{exp_it} \end{equation}
and hence
\begin{eqnarray*}
Z(\mathbf{b}_k)\ket{+_{-j}} &=& \frac{1}{\sqrt{d}} \sum_{l=0}^{d-1} \omega^{\alpha_{l+jk} + \frac{1}{2}j(j-1)k} \ket{l} \\
&=& \omega^{\frac{1}{2}j(j-1)k} \frac{1}{\sqrt{d}} \sum_{l=0}^{d-1} \omega^{\alpha_{l+jk}} \ket{l} \\
&=& \omega^{\frac{1}{2}j(j-1)k} \frac{1}{\sqrt{d}} \sum_{l=0}^{d-1} \omega^{\alpha_l} \ket{l-jk} \\
&=& \omega^{\frac{1}{2}j(j-1)k} \ket{\psi^k_{-jk}} 
\end{eqnarray*}
So we have that
\[ Z(\mathbf{b}_k)\ket{+_j} = \omega^{\frac{1}{2}j(j+1)k} \ket{\psi^k_{jk}} \]
and since the dimension here is prime, as $j$ runs over all values $0,\ldots, k-1$, so does $jk$. }
\vspace{10pt}
We can recover lemma \ref{EP} from this by the phase transformation $\ket{\psi^k_j} \mapsto \omega^{\frac{1}{2}j(jk^{-1}+1)}\ket{\psi^k_j}$, hence removing the phase that appears in the above lemma.

The eigenvectors of $ZX$ for $d=2$ are given by $\ket{\psi_\pm}= \frac{1}{\sqrt{2}}\left( \ket{0} \pm i \ket{1} \right)$. The above proof fails because (\ref{exp_it}) is only self-consistent when $d$ is odd: By substituting $j$ for $d$, then modulo $d$, we obtain that $\alpha_l \equiv \alpha_{l+dk} + dl + \frac{1}{2}d(d-1)k \equiv \alpha_{l} + \frac{1}{2}d(d-1)k \mod d$ i.e. we require $\frac{1}{2}d(d-1)k \equiv 0 \mod d$, which is true if and only if $d$ is odd. However, lemma \ref{EP} still holds: let $\mathbf{b} = (0,\pi)$. Then $Z(\mathbf{b}) = \oper{0}{0} + i \oper{1}{1}$, and it is easy to verify that
\[ Z(\mathbf{b}) \ket{+_0} = \ket{\psi_+}; \quad Z(\mathbf{b}) \ket{+_1} = \ket{\psi_-}. \]

\section{The Clifford group in prime dimension} \label{app_c}

The sole result of this appendix is a result of Gottesman \cite{GP}:
\vspace{10pt}
\begin{lem} The Clifford group $C_d$ (normaliser of the generalised Pauli group $\mathcal{P}_d$) is, when $d$ is a prime, generated by the operators $Z, X, F, S_c$ and $P$ defined by $P: \ket{j} \to \omega^{j(j+1)/2}\ket{j}$. \end{lem}
\vspace{10pt}
\proof{ The one-qudit Pauli group consists of elements of the form $\omega^a X^b Z^c$. Let us (for now) ignore the preceding powers of $\omega$. Then, under conjugation by a Clifford group operation (i.e. for $U \in C_d$, the transformation is the conjugation $P \mapsto UPU^\dagger$), we must have that (up to phases) $X \mapsto X^iZ^j$ and $Z \mapsto X^kZ^l$.
For $A,B \in \mathcal{P}_d$, define $\alpha(A,B)$ by $AB = \omega^{\alpha(A,B)}BA$. Under conjugation by a Clifford operation, this commutation relation is preserved and so we must have that $\alpha(X^iZ^j,X^kZ^l)=\alpha(X,Z)=1$. Furthemore,
\begin{eqnarray*}
\alpha(X^iZ^j,X^kZ^l) &=& \alpha(Z^j, X^k) + \alpha(X^i, Z^l) \\
&=& jk\alpha(Z,X) + il\alpha(X,Z) =il-jk 
\end{eqnarray*}
and so a general Clifford operation is an element of $\mathcal{P}_d$ (to choose the phases for the images of $X$ and $Z$) times some operation from some class of operations labelled by $(i,j,k,l)$, with $il-jk=1$.
We know that the Pauli operators are in the Clifford group, and so we can ignore any phases as the Pauli operators can correct for these. We split up considering the full set of operations into two groups. First let us assume that $i \neq 0$. It can be shown via elementary methods that if we define $C(i,m,n) = S_i P^m Q^n$ (where $Q=FPF^\dagger$), then, under conjugation by $C(i,m,n)$,
\[ X \mapsto X^i Z^{-i^{-1}m}; \quad Z \mapsto X^{in}Z^{i^{-1}(1-mn)} \]
(note that since $i \neq 0$, $i$ always has an inverse). By choosing $m=-ij$ and $n=i^{-1}k$, we find that $X \mapsto X^iZ^j$ and $Z \mapsto X^kZ^l$, with $l=i^{-1}(1+jk)$ as required. This gives us all Clifford operations for $i \neq 0$. 
For $i=0$, we must then have $jk=1$, and so $j \neq 0$. We wish to implement the operation $X \mapsto Z^j$ and $Z \mapsto X^kZ^l$. We can do this by first performing $X \mapsto X^{-j}$ and $Z \mapsto Z^kX^{-l}$ (which can be done from the calculations above) and then conjugating by $F$ to implement the required transformation. This gives us all Cifford group operations. }

The above proof does not work for non-prime dimensions because of the need for inverses of all non-zero elements in $\mathbb{Z}_d$, which only exist if $d$ is prime.

\end{document}